\documentclass[a4paper,12pt,preprint]{article}
\usepackage{amsmath}
\usepackage{latexsym,amssymb}
\usepackage[mathscr]{eucal}
\usepackage{amsthm,amsxtra,amscd,upref}
\usepackage{layout,bm,dcolumn}
\usepackage{graphicx,color}
\usepackage{calc}
\usepackage{natbib}
\usepackage{setspace}

\newcommand{\DST}[1]{{\ensuremath{\displaystyle{#1}}}}
\newcommand{\DSF}[2]{{\ensuremath{\displaystyle{\frac{\DST{#1}}{\DST{#2}}}}}}
\newcommand{\vk}[1]{{\ensuremath{\underline{\bm{#1}}}}}
\newcommand{\vr}[0]{{\ensuremath{\vk{r}}}}
\newcommand{\vQ}[0]{{\ensuremath{\vk{Q}}}}

\newcommand{\lrb}[1]{{\ensuremath{\left({#1}\right)}}}

\newcommand{\lrv}[1]{{\ensuremath{\left|{#1}\right|}}}

\newcommand{\DD}[1]{{\ensuremath{\mathrm{d}{#1}\, }}}
\newcommand{\DDD}[2]{{\ensuremath{\mathrm{d}^{#1}{#2}\, }}}
\newcommand{\sinc}[0]{{\ensuremath{\mathrm{sinc}}}}

\newcommand{\IMA}{{\ensuremath{\mathrm{i}}}}
\newcommand{\EE}{{\ensuremath{\mathrm{e}}}}

\newcommand{\citeb}[1]{{(\cite{#1})}}

\begin{document}
                
\title{{\textbf{A dilute gold nanoparticles suspension as SAXS standard for absolute scale using an extended Guinier approximation}}}
\author{
\begin{minipage}[t]{0.8\textwidth}
{
Ahmed S. A. Mohammed$^{a,b,c}$, Agnese Carino$^{d}$, 
Andrea Testino$^{d}$, Mohamed Reza Andalibi$^{d}$, 
Antonio Cervellino$^{a,*}$\\
\ \\
{\small{$^a$~Paul Scherrer Institute (PSI), Swiss Light Source (SLS), CH-5232 Villigen PSI, Switzerland\\ 
$^b$~Physik-Institut der Universit\"{a}t Z\"{u}rich, Winterthurerstrasse 190, CH-8057 Z\"{u}rich, Switzerland\\
$^c$~Physics Department, Faculty of Science, Fayoum University, 63514 Fayoum, {Egypt}\\
$^d$~Paul Scherrer Institute (PSI), ENE, CH-5232 Villigen PSI, {Switzerland}\\
\ \\
$^*$~{\small{\emph{Corresponding author. Email: antonio.cervellino@psi.ch}}}
}}
}\end{minipage}}
\date{\today}

\maketitle                        
\doublespacing

\begin{abstract}
In this article, a practical procedure for absolute intensity calibration for SAXS studies on liquid microjets is established, 
using a gold nanoparticle suspension as standard so that the intercept at $Q=0$ of the SAXS scattering curve would provide 
a scaling reference. In order to get the most precise extrapolation at $Q=0$, 
we used an extension to the Guinier approximation, with a second-order term in the fit that adapts to a larger $Q$-range. 
\end{abstract}

\section{Introduction}

We were conducting a number of SAXS experiments on liquid jets 
requiring an absolute scale determination in order to estimate concentrations of very dilute samples in a dynamical state, 
whereas offline measurements could not yield the desired results. 
Moreover, experiments were performed on a beamline that is not a dedicated SAXS instrument, 
where the accent is on flexibility of operation. Therefore instrument precalibration is excluded as too costly in terms of time and resources. 
Instead, we use SAXS standard of which everything is known and from which a predictable SAXS signal could be obtained 
in exactly the same experimental configuration, in order to derive the absolute scale constant from the extrapolation of the 
SAXS intensity to $Q\rightarrow0$. Such a suitable standard was identified as a dilute suspension of known concentration 
(200 mg/L) of practically monodisperse oleate-capped Au nanoparticles, whose diameter was estimated to be 18~nm. 
The extrapolation to $Q\rightarrow0$ was achieved using an extended Guinier method. 
In fact, in order to use an extended data range, the classical Guinier approximation was extended to one more 
term (\emph{i.e.}, to the fourth order in $Q$). 

\section{Liquid microjet setup}

A free liquid jet setup was developed to collect \emph{in-situ} synchrotron small and wide angle X-ray scattering data 
\citeb{mohammed2017situ}. Among different applications, this approach enables the study of the early stage 
of solids formation (nucleation). Such a mechanism is delicate and it is prone to be influenced by 
many factors \citeb{carino2017thermodynamic,marmiroli2009free}; in this fields, \emph{in-situ} techniques 
are well-established and are able to provide reliable experimental results \citeb{haberkorn2003early}. 
A successful SAXS measurements setup was demonstrated by Schmidt and coworkers \citeb{schmidt2010accessing} for ZnS precipitation.
Here the aim is to study by \emph{in-situ} SAXS measurements very diluted inorganic systems containing entities with an 
electron density slightly higher than that of the solvent. 
Our study was specifically designed to investigate the biominarals formation pathway, 
such as calcium carbonate and calcium phosphate, under controlled temperature, pH, and saturation conditions.
The pulsation-free micrometric-size horizontal reacting liquid jet setup is composed by four HPLC pumps (each of them equipped 
with a pulsation damper system and a high precision Coriolis mass flowmeter), a micromixing system, a delay loop, and a catcher. 
The micromixer manifolds is equipped with 5 inputs and one output. Four inputs are connected to the solutions contributing 
to the precipitation reaction whereas the fifth input was used for cleaning purposes and connected to 
an additional pump delivering 10 wt.$\%$  acetic acid. 
A delay loop (which consists of a teflon tube of a certain length and internal diameter) can be connected the micromixer output. 
This tube determines a delay time between the mixing point and the measuring point, where the free liquid jet is interrogated by the 
X-ray beam. A double-walled water-jacketed tubing system is used to thermostat the delay tube and the chemical delivery lines 
before the mixing point. 
A nozzle of different materials and sizes may be connected after the delay loop and used to generate the liquid jet. 
In this study, a stainless steel capillary with internal diameter of 250 $\mu$m and an overall flow rate of 8 ml min$^{-1}$ was used. 
The main advantage of this approach is that the solid formation mechanisms can be followed while the liquid jet is continuously 
renovated, at a defined timeframe after the mixing point, therefore in static conditions with respect to the entities development, 
produced by the reaction, and with negligible perturbations of jet probed by the beam (negligible beam damage). Since 
the amount and electron density of the matter of interest is very low, an accurate experimental data collection needs to be carried 
out and data accumulated for a relatively long time at each experimental condition (static mode). Moreover, the absolute scale 
constant need to be evaluated. To this end, a well characterized diluted standard solution containing gold nanoparticles was used applying 
exactly the same experimental configuration of the precipitation experiments.

\section{Standard preparation and characterization}

The gold nanoparticle (GNP) suspension was synthesised using the citrate route, also called Turkevich method(\emph{i.e.}, by reducing the  tetrachloroauric acid (TCAA) precursors with trisodium citrate (Na$_3$cit) in water). The preparation was done using a segmented flow tubular reactor, SFTR \citeb{jongen2003development,donnet2000new,aimable2011precipitation}. The SFTR allows monitoring accurately the ratio of the chemicals, the temperature, and the aging time. The segmentation is obtained using a secondary immiscible fluid which also prevents fouling. The precursor solutions are mixed at room temperature and segmented into the tubular reactor as a sequence of droplets in which the reaction occurs triggered by the fast temperature rising of the tubular reactor. The obtained nanoparticle size distribution is narrow and reproducible. This reactor consents a continuous production up to 5 g of solid per day.
The conditions used to prepare the GNP suspension were similar to the ones reported in a previous work \citeb{carino2016continuous}. The synthesis was done at 95 $^\circ$C using a Na$_3$cit to TCAA molar ratio equal to 7.9, an ageing time of 5 min, and the chemicals flow rate was 1 L h$^{-1}$. The final GNP concentration was of about 0.2 g L$^{-1}$ (measured by ICP-MS). The suspension presented a monomodal distribution of GNP with an average number diameter of 18 nm and a standard deviation $<$ 8 $\%$ (measured by TEM imaging and dynamic light scattering).

\section{Data collection}

Synchrotron Small-Angle X-rays Scattering (SAXS) measurements were carried out at the Material Science beamline (X04SA$-$MS) of the Swiss Light Source (SLS) at PSI \citeb{willmott2013materials}. This synchrotron station is built mainly for WAXS powder diffraction measurements but it also has some SAXS capabilities. The liquid jet, horizontal and orthogonal to the X-ray beam, was mounted on a double micrometric translation stage and optically centered with respect to the diffractometer circle by a high-resolution camera. 
The X-ray beam was set at 9.5 keV (corresponding to a wavelength $\lambda=1.305  \AA$) where the X-ray flux is maximal. 
Data were collected with the Mythen II detector system \citeb{bergamaschi2010mythen}, that with its 0.0036  $deg$ step, 
has a sufficient resolution for SAXS on this system with a minimum accessible $2 \theta$  scattering angle of 
$\approx$ 0.18 $deg$, corresponding to 
a minimum accessible momentum transfer of $Q=4\pi\sin(\theta)/\lambda=0.015$~\AA$^{-1}$, 
Scattering patterns were collected with reasonable acquisition times (20 min). Before the analysis, 
the experimental data were processed, corrected and rebinned to an uniform step $Q$ grid, each data point being a 
triple $Q$-intensity-standard deviation.

\subsection{Extended Guinier approximation}
\label{exguapp}

The Guinier approximation is a universal method \citeb{guiner1955small}
to derive some fundamental parameters 
from SAXS data exploiting some universal behaviour for $Q\rightarrow 0$ of the scattered intensity 
for diluted systems. 
If $I(Q)$ is the sample's scattered intensity, already subtracted of background terms and 
corrected for all instrumental aberrations, the Guinier approximation 
\[
I(Q)\approx A_0\EE^{-bQ^2}
\]
where $A_0=I(Q=0)$ and $b=R_g^2/3$ is related to the sample (appropriately averaged) 
gyration radius (see Sec.~\ref{RMO}, Eq.~\ref{r2mo}). This form 
is valid because $I(Q)$ is a smooth positive even function with a maximum at $Q=0$ and 
monotonically decreasing for sufficiently small $Q$.  
It is usually recast as its MacLaurin expansion
\[
\log\lrb{I(Q)}\approx \log\lrb{A_0}-bQ^2
\]
A line fit of $\log\lrb{I(Q)}$ \emph{vs.} $Q^2$ yields then easily the values of $A_0$ and $b$. 
The range of validity of this approximation is however restricted to about
\[
Q < \DSF{\pi}{D}
\]
where $D$ is the scattering particle's diameter (in the case of a sphere). 

A straightforward extension of this approximation is surely possible, whereas we suppose that $I(Q)$ at small $Q$ can be 
described in the form
\begin{equation}
I(Q)\approx \EE^{a-bQ^2+CQ^4}
\end{equation}
that is, with a further quartic term in the exponent. 
Again we do the Mac Laurin expansion
\begin{equation}
\log\lrb{I(Q)}\approx \log\lrb{A_0}-bQ^2+cQ^4, \qquad c\equiv C+\DSF{b^2}{2}
\label{xtend}
\end{equation}
where a second order polynomial in $Q^2$ replaces the straight line. A more technical derivation is given in Appendix~\ref{XX}.

\subsection{Determination of the scale factor}

We employ a water suspension of spherical Au nanoparticles of known concentration (200 mg/l) and 
known, practically unimodal size (18-19 nm) as standard to determine the absolute scale factor $k$. 
Such a suspension can be measured with the liquid jet with exactly the same setup as the investigated 
samples in this project. In fact, the observed intensity per unit time (after subtracting the background, 
including that of the liquid, measured separately) is
\begin{equation}
\label{eq:sc1}
I(Q)=k N (\Delta \rho)^2 V_p^2 \Bigg(3 \frac {\sin(QR)-QR \cos(QR)}{(QR)^3}\Bigg)^2
\end {equation}
where $k$ is a constant pertaining to the experimental setup, $N$ is the number of illuminated particles, $\Delta \rho$ 
is the particles' scattering contrast (the difference in scattering length density with respect to the average; 
hereafter we use electron units, 
where the classical electron radius $r_e=1$, 
so $\Delta \rho$ has units of an inverse volume), $V_p$ is the particle volume and the term within brackets is the analytical 
expression of the scattered intensity of a sphere. 
\begin{figure}
\begin{center}
\includegraphics[width=1\textwidth]{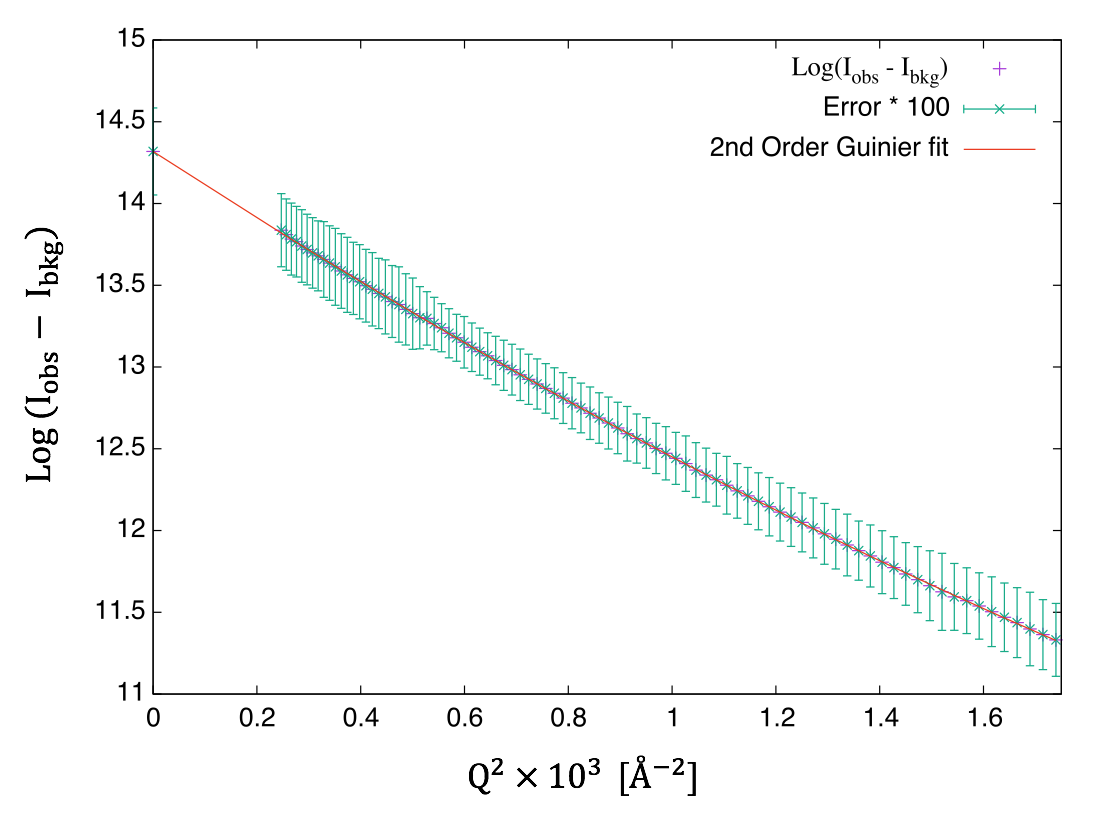}
\caption{Guinier fit with second-order term added in order to exploit a larger Q-range and obtain a good estimate of the intercept, 
as needed for the absolute scale factor estimation.}
\label{fig:fg1}
\end{center}
\end{figure}

The SAXS patterns were analyzed using the extended Guinier approximation of Eq.~\ref{xtend}, discussed in Sec.~\ref{exguapp}. 
The parabolic fit is shown in Fig. ~\ref {fig:fg1}. 
As our $Q$-range is 
contains relatively few points within the strict limits of validity of the standard Guinier approximation, 
the quadratic term in the fit allowed us to include an about three times wider $Q$-range, 
up to about $Q \approx 6/D$. 
The significance of the second order term is immaterial within the scope of this paper. 
Surely it helps gaining a more precise estimate of the other two parameters, especially $A_0$. 
The extended Guinier fit yielded a gyration radius of $7.8799 \pm 0.0116$~$nm$ 
corresponding - in the case of perfectly monodisperse spherical particles - to a particle 
diameter of $20.35 \pm 0.03$~nm, consistent (within the limits of the assumptions above) with an estimated diameter of 18 $nm$. 
More importantly, the extrapolated forward intensity was found to be
\[
A_0=I(Q=0) =
1.6534\times 10^6    \ \pm\    4.4\times 10^{-3}
\quad\text{counts per second}
\]
The average electron density of the suspension can be approximated with that of pure water, given the low concentration. 
For the Au NPs the density of bulk gold will be used. 
Therefore, evaluating the electron densities yields
$\rho_{\mathrm{H_2O}}=3.34572\times 10^{23}\quad {\mathrm{cm^{-3}}};$ 
$\rho_{\mathrm{Au}}=4.66101\times 10^{24}\quad {\mathrm{cm^{-3}}};$
$\Delta\rho=4.32644\times 10^{24}\quad {\mathrm{cm^{-3}}}$
From the given concentration and the Guinier diameter, we can assess that the number of Au NPs per cubic cm$^3$ is
\(
n=2.3499\times10^{12}\quad{\mathrm{cm^{-3}}}
\)
and as the illuminated liquid volume was calculated as
\(
V=1.96350\times10^{-4} \quad{\mathrm{cm^{3}}}
\) 
so we have a total of $N=nV=4.614\times10^{8}$ illuminated NPs. The particle volume, 
from the estimated diameter of 20.346~nm, 
results to
$V_p=4.4098\times10^{-18}
\ {\mathrm{cm^{3}}}$
At $Q=0$, the term in brackets in equation ~\ref{eq:sc1} is 1, so we can now solve for the sought-after scale factor:
\begin{equation}
k=\DSF{I(Q=0)}{N\lrb{\Delta\rho}^2V_p^2}= 
1.0114\times10^{-17}
\label{kval}
\end{equation}

\section{Conclusions}

We have demonstrated an external standard method to evaluate the scale factor for absolute intensity in SAXS experiments 
involving a liquid jet containing nanoparticles. The standard for liquid jet-type experiments is a suspension of as much as possible 
monodisperse nanoparticles of known concentration, high contrast, spherical shape and a known size that is similar
to that of the samples under investigation. An extended Guinier approximation has been introduced, in order to improve 
the fit and consequently to achieve a more precise estimate of the extrapolated intensity at $Q=0$. 
The method is simple, cheap, rapid (as it requires only one additional scattering measurement) and easily applicable to any 
beamline or instrument where a liquid jet device is installed. This also removes the need to perform a much tougher calibration 
of the instrument when absolute scale data are desired, while preserving also the same flexibility of operation of relative-scale instruments.


\appendix

\section{Theory}\label{XX}

The Guinier approximation can be more rigorously derived
by using its universal expression given by the Debye scattering equation (free from interparticle interference effects, 
as appropriate for diluted systems). 
The Debye scattering equation \citeb{debyetotalscatt} yields the orientation averaged differential cross section 
directly from the atomic scattering density. 
Note that as we anyway consider a very small $Q$ range, we can assume that atomic scattering factors (scattering lengths) are constant 
and independent from $Q$ also for X-rays, slightly simplifying the discussion. 

The atomic scattering density of a particle containing $N_a$ atoms 
that have scattering lengths (or scattering factors) $b_j$ and are located at positions $\vr_j$ for $j=1,\ldots,N_a$ is represented as
\[
\rho\lrb{\vr} = \mathop{\sum}_{j=1}^{N_a}b_j\,\delta\lrb{\vr-\vr_j}
\]
As the origin is arbitrary, we refer the atomic coordinates to the scattering center of mass, so that
\begin{equation}
 \mathop{\sum}_{j=1}^{N_a}b_j\vr_j \equiv 0
 \label{cmsc}
\end{equation}
We also call $B_1$, $B_2$ the sum of scattering lengths and that of their squares:
\begin{equation}
B_1\equiv \mathop{\sum}_{j=1}^{N_a}b_j
\,;\qquad
B_2\equiv \mathop{\sum}_{j=1}^{N_a}b_j^2
\label{EB2}
\end{equation}

\subsection{Radial moments}\label{RMO}

The radial moments of a density are defined as
\[
M_n=\DSF{\int\DDD{3}{\vr}\rho\lrb{\vr}\vr^n}{\int\DDD{3}{\vr}\rho\lrb{\vr}}
\]
where $\vr^{2m}\equiv(\vr\cdot\vr)^m=r^{2m}$ for even $n$, 
$\vr^{2m+1}\equiv(\vr\cdot\vr)^m\vr=r^{2m}\vr$ for odd $n$. 
Thanks to Eq.~(\ref{cmsc}), the odd moments are all zero. 
We can now evaluate the even radial moments of this density:
\begin{equation}
M_{2n}=\DSF{\int\DDD{3}{\vr}\rho\lrb{\vr}r^{2n}}{\int\DDD{3}{\vr}\rho\lrb{\vr}}=
\DSF{1}{B_1}{\mathop{\sum}_{j=1}^{N_a}b_jr_j^{2n}}
\label{r2mo}
\end{equation}
In particular, $M_2$ is called the squared gyration radius 
\[
r_g^2=M_2
\]

The differential cross section as a function of the transferred momentum $\vQ$ (where $Q=4\pi \sin\theta/\lambda$) is
\begin{equation}
I\lrb{\vQ}=\mathop{\sum}_{j,k=1}^{N_a}b_jb_k\,\EE^{\IMA\vQ\cdot\lrb{\vr_j-\vr_k}}=
B_2+2\mathop{\sum}_{j>k=1}^{N_a}b_jb_k\,\EE^{\IMA\vQ\cdot\lrb{\vr_j-\vr_k}}
\end{equation}
Here in evidence (first term on the RHS) is the self-scattering $B_2$ (\emph{cf.} Eq. \ref{EB2}). 
Making the orientation average over all possible $\vQ$ directions, 
we obtain the Debye scattering equation (with $\sinc(x)\equiv\sin(x)/x$)
\begin{equation}
I(Q)=\mathop{\sum}_{j,k=1}^{N_a}b_jb_k\sinc\lrb{Q\lrv{\vr_j-\vr_k}}=
B_2+2\mathop{\sum}_{j>k=1}^{N_a}b_jb_k\,\sinc\lrb{Qd_{jk}}\label{DSE}
\end{equation}
where $d_{jk}\equiv\lrv{\vr_j-\vr_k}$.
Now we consider only the structural term $I_s(Q)=I(Q)-B_2$.
The pair correlation function, that is the sinc transform of $I_s(Q)$, is
\[
f_s(r)=\int_0^{+\infty}4\pi Q^2\DD{Q}\sinc\lrb{Qr}I_s(Q)= 
2\mathop{\sum}_{j>k=1}^{N_a}\DSF{b_jb_k}{4\pi rd_{jk}}\,\delta\lrb{r-d_{jk}}
\]
We want to calculate\\
\begin{equation}
Z_2=\int_0^{+\infty}4\pi r^2\DD{r}f_s(r)r^2 = 
2\mathop{\sum}_{j>k=1}^{N_a}{b_jb_k}{d_{jk}^2}=
\mathop{\sum}_{j,k=1}^{N_a}{b_jb_k}\lrb{r_{j}^2+r_k^2-2\vr_j\cdot\vr_k}
\label{RGZ}
\end{equation}\\
where we have expanded the distance squared and re-added the diagonal terms as they are 0 anyway.
Now we split the sum in 3 parts, each with one of the terms in brackets, and reorder:
\begin{eqnarray}
\mathop{\sum}_{j,k=1}^{N_a}{b_jb_k}r_{j}^2=\mathop{\sum}_{j,k=1}^{N_a}{b_jb_k}r_{k}^2&=&B_1 \mathop{\sum}_{j=1}^{N_a}{b_j}r_{j}^2= B_1^2 r_g^2 \\
-2\mathop{\sum}_{j,k=1}^{N_a}{b_jb_k}\vr_j\cdot\vr_k&=&-2\lrb{
\mathop{\sum}_{k=1}^{N_a}b_k\vr_k 
} \cdot\lrb{\mathop{\sum}_{j=1}^{N_a}b_j\vr_j} =0
\end{eqnarray}
the last is a consequence of of Eq.~(\ref{cmsc}).
Therefore $Z_2=2B_1^2r_g^2$ and the gyration radius can be evaluated also directly from the pair correlation function, as
\[
r_g^2=\DSF{Z_2}{2B_1^2}
\]
using Eqs.~(\ref{EB2},\ref{RGZ}). 

The Guinier approximation can be now derived from Eq.~(\ref{DSE}). 
There we can expand $I(Q)$ for small $Q$ in a Mac Laurin series: \\
\begin{equation}
I(Q)=\mathop{\sum}_{j,k=1}^{N_a}b_jb_k -\DSF{Q^2}{6}\mathop{\sum}_{j,k=1}^{N_a}b_jb_kd_{jk}^2 
+\DSF{Q^4}{120}\mathop{\sum}_{j,k=1}^{N_a}b_jb_kd_{jk}^4 
-\DSF{Q^6}{5040}\mathop{\sum}_{j,k=1}^{N_a}b_jb_kd_{jk}^6 
+\ldots
\end{equation}\\
Being $I(Q)$ a positive, continuous and even function, 
its logarithm can also be expanded in a MacLaurin series 
that will contain only even powers of $Q$. 
Therefore, if we equate 
\[
I(Q)=\exp\lrb{\mathop{\sum}_{m=1}^{+\infty}
a_mQ^{2m}
}
\]
 and expand it also in a MacLaurin series, 
we can then equate coefficients of equal powers of $Q$. The first 3 terms (up to $Q^4$) yield three equations:
\begin{eqnarray}
 \EE^{a_0} &=& \mathop{\sum}_{j,k=1}^{N_a}b_jb_k = B_1^2\\
a_1\EE^{a_0} &=&  -\DSF{1}{6}\mathop{\sum}_{j,k=1}^{N_a}b_jb_kd_{jk}^2=-\DSF{1}{3}B_1^2r_g^2 
\\\DSF{1}{2}\lrb{a_1^2+2a_2}\EE^{a_0} &=& \DSF{1}{120}\mathop{\sum}_{j,k=1}^{N_a}b_jb_kd_{jk}^4
\end{eqnarray}
and so on. Solving the first 2 equations, 
we have
\[
a_0=2\log\lrb{B_1}; \qquad
a_1=-\DSF{1}{3}B_1^2r_g^2
\]
as in the classical Guinier approximation. 
From the third equation one could derive an expression for a higher radial moments $M_4$.


\begin{thebibliography}{13}%
\makeatletter
\providecommand \@ifxundefined [1]{%
 \@ifx{#1\undefined}
}%
\providecommand \@ifnum [1]{%
 \ifnum #1\expandafter \@firstoftwo
 \else \expandafter \@secondoftwo
 \fi
}%
\providecommand \@ifx [1]{%
 \ifx #1\expandafter \@firstoftwo
 \else \expandafter \@secondoftwo
 \fi
}%
\providecommand \natexlab [1]{#1}%
\providecommand \enquote  [1]{``#1''}%
\providecommand \bibnamefont  [1]{#1}%
\providecommand \bibfnamefont [1]{#1}%
\providecommand \citenamefont [1]{#1}%
\providecommand \href@noop [0]{\@secondoftwo}%
\providecommand \href [0]{\begingroup \@sanitize@url \@href}%
\providecommand \@href[1]{\@@startlink{#1}\@@href}%
\providecommand \@@href[1]{\endgroup#1\@@endlink}%
\providecommand \@sanitize@url [0]{\catcode `\\12\catcode `\$12\catcode
  `\&12\catcode `\#12\catcode `\^12\catcode `\_12\catcode `\%12\relax}%
\providecommand \@@startlink[1]{}%
\providecommand \@@endlink[0]{}%
\providecommand \url  [0]{\begingroup\@sanitize@url \@url }%
\providecommand \@url [1]{\endgroup\@href {#1}{\urlprefix }}%
\providecommand \urlprefix  [0]{URL }%
\providecommand \Eprint [0]{\href }%
\providecommand \doibase [0]{http://dx.doi.org/}%
\providecommand \selectlanguage [0]{\@gobble}%
\providecommand \bibinfo  [0]{\@secondoftwo}%
\providecommand \bibfield  [0]{\@secondoftwo}%
\providecommand \translation [1]{[#1]}%
\providecommand \BibitemOpen [0]{}%
\providecommand \bibitemStop [0]{}%
\providecommand \bibitemNoStop [0]{.\EOS\space}%
\providecommand \EOS [0]{\spacefactor3000\relax}%
\providecommand \BibitemShut  [1]{\csname bibitem#1\endcsname}%
\let\auto@bib@innerbib\@empty
\bibitem [{\citenamefont {Mohammed}\ \emph {et~al.}(2017)\citenamefont
  {Mohammed}, \citenamefont {Cervellino}, \citenamefont {Testino},\ and\
  \citenamefont {Carino}}]{mohammed2017situ}%
  \BibitemOpen
  \bibfield  {author} {\bibinfo {author} {\bibfnamefont {A.~S.}\ \bibnamefont
  {Mohammed}}, \bibinfo {author} {\bibfnamefont {A.}~\bibnamefont
  {Cervellino}}, \bibinfo {author} {\bibfnamefont {A.}~\bibnamefont {Testino}},
  \ and\ \bibinfo {author} {\bibfnamefont {A.}~\bibnamefont {Carino}},\
  }\href@noop {} {\bibfield  {journal} {\bibinfo  {journal} {Acta Crystallogr.
  A Suppl.}\ }\textbf {\bibinfo {volume} {70}},\ \bibinfo {pages} {C315}
  (\bibinfo {year} {2017})}\BibitemShut {NoStop}%
\bibitem [{\citenamefont {Carino}\ \emph {et~al.}(2017)\citenamefont {Carino},
  \citenamefont {Testino}, \citenamefont {Andalibi}, \citenamefont {Pilger},
  \citenamefont {Bowen},\ and\ \citenamefont
  {Ludwig}}]{carino2017thermodynamic}%
  \BibitemOpen
  \bibfield  {author} {\bibinfo {author} {\bibfnamefont {A.}~\bibnamefont
  {Carino}}, \bibinfo {author} {\bibfnamefont {A.}~\bibnamefont {Testino}},
  \bibinfo {author} {\bibfnamefont {M.~R.}\ \bibnamefont {Andalibi}}, \bibinfo
  {author} {\bibfnamefont {F.}~\bibnamefont {Pilger}}, \bibinfo {author}
  {\bibfnamefont {P.}~\bibnamefont {Bowen}}, \ and\ \bibinfo {author}
  {\bibfnamefont {C.}~\bibnamefont {Ludwig}},\ }\href@noop {} {\bibfield
  {journal} {\bibinfo  {journal} {Crystal Growth \& Design}\ }\textbf {\bibinfo
  {volume} {17}},\ \bibinfo {pages} {2006} (\bibinfo {year}
  {2017})}\BibitemShut {NoStop}%
\bibitem [{\citenamefont {Marmiroli}\ \emph {et~al.}(2009)\citenamefont
  {Marmiroli}, \citenamefont {Grenci}, \citenamefont {Cacho-Nerin},
  \citenamefont {Sartori}, \citenamefont {Ferrari}, \citenamefont {Laggner},
  \citenamefont {Businaro},\ and\ \citenamefont
  {Amenitsch}}]{marmiroli2009free}%
  \BibitemOpen
  \bibfield  {author} {\bibinfo {author} {\bibfnamefont {B.}~\bibnamefont
  {Marmiroli}}, \bibinfo {author} {\bibfnamefont {G.}~\bibnamefont {Grenci}},
  \bibinfo {author} {\bibfnamefont {F.}~\bibnamefont {Cacho-Nerin}}, \bibinfo
  {author} {\bibfnamefont {B.}~\bibnamefont {Sartori}}, \bibinfo {author}
  {\bibfnamefont {E.}~\bibnamefont {Ferrari}}, \bibinfo {author} {\bibfnamefont
  {P.}~\bibnamefont {Laggner}}, \bibinfo {author} {\bibfnamefont
  {L.}~\bibnamefont {Businaro}}, \ and\ \bibinfo {author} {\bibfnamefont
  {H.}~\bibnamefont {Amenitsch}},\ }\href@noop {} {\bibfield  {journal}
  {\bibinfo  {journal} {Lab on a Chip}\ }\textbf {\bibinfo {volume} {9}},\
  \bibinfo {pages} {2063} (\bibinfo {year} {2009})}\BibitemShut {NoStop}%
\bibitem [{\citenamefont {Haberkorn}\ \emph {et~al.}(2003)\citenamefont
  {Haberkorn}, \citenamefont {Franke}, \citenamefont {Frechen}, \citenamefont
  {Goesele},\ and\ \citenamefont {Rieger}}]{haberkorn2003early}%
  \BibitemOpen
  \bibfield  {author} {\bibinfo {author} {\bibfnamefont {H.}~\bibnamefont
  {Haberkorn}}, \bibinfo {author} {\bibfnamefont {D.}~\bibnamefont {Franke}},
  \bibinfo {author} {\bibfnamefont {T.}~\bibnamefont {Frechen}}, \bibinfo
  {author} {\bibfnamefont {W.}~\bibnamefont {Goesele}}, \ and\ \bibinfo
  {author} {\bibfnamefont {J.}~\bibnamefont {Rieger}},\ }\href@noop {}
  {\bibfield  {journal} {\bibinfo  {journal} {Journal of Colloid and Interface
  Science}\ }\textbf {\bibinfo {volume} {259}},\ \bibinfo {pages} {112}
  (\bibinfo {year} {2003})}\BibitemShut {NoStop}%
\bibitem [{\citenamefont {Schmidt}\ \emph {et~al.}(2010)\citenamefont
  {Schmidt}, \citenamefont {Bussian}, \citenamefont {LindÈn}, \citenamefont
  {Amenitsch}, \citenamefont {Agren}, \citenamefont {Tiemann},\ and\
  \citenamefont {Sch?th}}]{schmidt2010accessing}%
  \BibitemOpen
  \bibfield  {author} {\bibinfo {author} {\bibfnamefont {W.}~\bibnamefont
  {Schmidt}}, \bibinfo {author} {\bibfnamefont {P.}~\bibnamefont {Bussian}},
  \bibinfo {author} {\bibfnamefont {M.}~\bibnamefont {LindÈn}}, \bibinfo
  {author} {\bibfnamefont {H.}~\bibnamefont {Amenitsch}}, \bibinfo {author}
  {\bibfnamefont {P.}~\bibnamefont {Agren}}, \bibinfo {author} {\bibfnamefont
  {M.}~\bibnamefont {Tiemann}}, \ and\ \bibinfo {author} {\bibfnamefont
  {F.}~\bibnamefont {Sch?th}},\ }\href@noop {} {\bibfield  {journal} {\bibinfo
  {journal} {Journal of the American Chemical Society}\ }\textbf {\bibinfo
  {volume} {132}},\ \bibinfo {pages} {6822} (\bibinfo {year}
  {2010})}\BibitemShut {NoStop}%
\bibitem [{\citenamefont {Jongen}\ \emph {et~al.}(2003)\citenamefont {Jongen},
  \citenamefont {Donnet}, \citenamefont {Bowen}, \citenamefont {Lema{\^\i}tre},
  \citenamefont {Hofmann}, \citenamefont {Schenk}, \citenamefont {Hofmann},
  \citenamefont {Aoun-Habbache}, \citenamefont {Guillemet-Fritsch},
  \citenamefont {Sarrias} \emph {et~al.}}]{jongen2003development}%
  \BibitemOpen
  \bibfield  {author} {\bibinfo {author} {\bibfnamefont {N.}~\bibnamefont
  {Jongen}}, \bibinfo {author} {\bibfnamefont {M.}~\bibnamefont {Donnet}},
  \bibinfo {author} {\bibfnamefont {P.}~\bibnamefont {Bowen}}, \bibinfo
  {author} {\bibfnamefont {J.}~\bibnamefont {Lema{\^\i}tre}}, \bibinfo {author}
  {\bibfnamefont {H.}~\bibnamefont {Hofmann}}, \bibinfo {author} {\bibfnamefont
  {R.}~\bibnamefont {Schenk}}, \bibinfo {author} {\bibfnamefont
  {C.}~\bibnamefont {Hofmann}}, \bibinfo {author} {\bibfnamefont
  {M.}~\bibnamefont {Aoun-Habbache}}, \bibinfo {author} {\bibfnamefont
  {S.}~\bibnamefont {Guillemet-Fritsch}}, \bibinfo {author} {\bibfnamefont
  {J.}~\bibnamefont {Sarrias}},  \emph {et~al.},\ }\href@noop {} {\bibfield
  {journal} {\bibinfo  {journal} {Chemical Engineering \& Technology:
  Industrial Chemistry-Plant Equipment-Process Engineering-Biotechnology}\
  }\textbf {\bibinfo {volume} {26}},\ \bibinfo {pages} {303} (\bibinfo {year}
  {2003})}\BibitemShut {NoStop}%
\bibitem [{\citenamefont {Donnet}\ \emph {et~al.}(2000)\citenamefont {Donnet},
  \citenamefont {Jongen}, \citenamefont {Lemaitre},\ and\ \citenamefont
  {Bowen}}]{donnet2000new}%
  \BibitemOpen
  \bibfield  {author} {\bibinfo {author} {\bibfnamefont {M.}~\bibnamefont
  {Donnet}}, \bibinfo {author} {\bibfnamefont {N.}~\bibnamefont {Jongen}},
  \bibinfo {author} {\bibfnamefont {J.}~\bibnamefont {Lemaitre}}, \ and\
  \bibinfo {author} {\bibfnamefont {P.}~\bibnamefont {Bowen}},\ }\href@noop {}
  {\bibfield  {journal} {\bibinfo  {journal} {Journal of Materials Science
  Letters}\ }\textbf {\bibinfo {volume} {19}},\ \bibinfo {pages} {749}
  (\bibinfo {year} {2000})}\BibitemShut {NoStop}%
\bibitem [{\citenamefont {Aimable}\ \emph {et~al.}(2011)\citenamefont
  {Aimable}, \citenamefont {Jongen}, \citenamefont {Testino}, \citenamefont
  {Donnet}, \citenamefont {Lema{\^\i}tre}, \citenamefont {Hofmann},\ and\
  \citenamefont {Bowen}}]{aimable2011precipitation}%
  \BibitemOpen
  \bibfield  {author} {\bibinfo {author} {\bibfnamefont {A.}~\bibnamefont
  {Aimable}}, \bibinfo {author} {\bibfnamefont {N.}~\bibnamefont {Jongen}},
  \bibinfo {author} {\bibfnamefont {A.}~\bibnamefont {Testino}}, \bibinfo
  {author} {\bibfnamefont {M.}~\bibnamefont {Donnet}}, \bibinfo {author}
  {\bibfnamefont {J.}~\bibnamefont {Lema{\^\i}tre}}, \bibinfo {author}
  {\bibfnamefont {H.}~\bibnamefont {Hofmann}}, \ and\ \bibinfo {author}
  {\bibfnamefont {P.}~\bibnamefont {Bowen}},\ }\href@noop {} {\bibfield
  {journal} {\bibinfo  {journal} {Chemical Engineering \& Technology}\ }\textbf
  {\bibinfo {volume} {34}},\ \bibinfo {pages} {344} (\bibinfo {year}
  {2011})}\BibitemShut {NoStop}%
\bibitem [{\citenamefont {Carino}\ \emph {et~al.}(2016)\citenamefont {Carino},
  \citenamefont {Walter}, \citenamefont {Testino},\ and\ \citenamefont
  {Hofmann}}]{carino2016continuous}%
  \BibitemOpen
  \bibfield  {author} {\bibinfo {author} {\bibfnamefont {A.}~\bibnamefont
  {Carino}}, \bibinfo {author} {\bibfnamefont {A.}~\bibnamefont {Walter}},
  \bibinfo {author} {\bibfnamefont {A.}~\bibnamefont {Testino}}, \ and\
  \bibinfo {author} {\bibfnamefont {H.}~\bibnamefont {Hofmann}},\ }\href@noop
  {} {\bibfield  {journal} {\bibinfo  {journal} {CHIMIA International Journal
  for Chemistry}\ }\textbf {\bibinfo {volume} {70}},\ \bibinfo {pages} {457}
  (\bibinfo {year} {2016})}\BibitemShut {NoStop}%
\bibitem [{\citenamefont {Willmott}\ \emph {et~al.}(2013)\citenamefont
  {Willmott}, \citenamefont {Meister}, \citenamefont {Leake}, \citenamefont
  {Lange}, \citenamefont {Bergamaschi}, \citenamefont {B{\"{o}}ge},
  \citenamefont {Calvi}, \citenamefont {Cancellieri}, \citenamefont {Casati},
  \citenamefont {Cervellino}, \citenamefont {Chen}, \citenamefont {David},
  \citenamefont {Flechsig}, \citenamefont {Gozzo}, \citenamefont {Henrich},
  \citenamefont {J{\"{a}}ggi-Spielmann}, \citenamefont {Jakob}, \citenamefont
  {Kalichava}, \citenamefont {Karvinen}, \citenamefont {Krempasky},
  \citenamefont {L{\"{u}}deke}, \citenamefont {L{\"{u}}scher}, \citenamefont
  {Maag}, \citenamefont {Quitmann}, \citenamefont {Reinle-Schmitt},
  \citenamefont {Schmidt}, \citenamefont {Schmitt}, \citenamefont {Streun},
  \citenamefont {Vartiainen}, \citenamefont {Vitins}, \citenamefont {Wang},\
  and\ \citenamefont {Wullschleger}}]{willmott2013materials}%
  \BibitemOpen
  \bibfield  {author} {\bibinfo {author} {\bibfnamefont {P.}~\bibnamefont
  {Willmott}}, \bibinfo {author} {\bibfnamefont {D.}~\bibnamefont {Meister}},
  \bibinfo {author} {\bibfnamefont {S.}~\bibnamefont {Leake}}, \bibinfo
  {author} {\bibfnamefont {M.}~\bibnamefont {Lange}}, \bibinfo {author}
  {\bibfnamefont {A.}~\bibnamefont {Bergamaschi}}, \bibinfo {author}
  {\bibfnamefont {M.}~\bibnamefont {B{\"{o}}ge}}, \bibinfo {author}
  {\bibfnamefont {M.}~\bibnamefont {Calvi}}, \bibinfo {author} {\bibfnamefont
  {C.}~\bibnamefont {Cancellieri}}, \bibinfo {author} {\bibfnamefont
  {N.}~\bibnamefont {Casati}}, \bibinfo {author} {\bibfnamefont
  {A.}~\bibnamefont {Cervellino}}, \bibinfo {author} {\bibfnamefont
  {Q.}~\bibnamefont {Chen}}, \bibinfo {author} {\bibfnamefont {C.}~\bibnamefont
  {David}}, \bibinfo {author} {\bibfnamefont {U.}~\bibnamefont {Flechsig}},
  \bibinfo {author} {\bibfnamefont {F.}~\bibnamefont {Gozzo}}, \bibinfo
  {author} {\bibfnamefont {B.}~\bibnamefont {Henrich}}, \bibinfo {author}
  {\bibfnamefont {S.}~\bibnamefont {J{\"{a}}ggi-Spielmann}}, \bibinfo {author}
  {\bibfnamefont {B.}~\bibnamefont {Jakob}}, \bibinfo {author} {\bibfnamefont
  {I.}~\bibnamefont {Kalichava}}, \bibinfo {author} {\bibfnamefont
  {P.}~\bibnamefont {Karvinen}}, \bibinfo {author} {\bibfnamefont
  {J.}~\bibnamefont {Krempasky}}, \bibinfo {author} {\bibfnamefont
  {A.}~\bibnamefont {L{\"{u}}deke}}, \bibinfo {author} {\bibfnamefont
  {R.}~\bibnamefont {L{\"{u}}scher}}, \bibinfo {author} {\bibfnamefont
  {S.}~\bibnamefont {Maag}}, \bibinfo {author} {\bibfnamefont {C.}~\bibnamefont
  {Quitmann}}, \bibinfo {author} {\bibfnamefont {M.}~\bibnamefont
  {Reinle-Schmitt}}, \bibinfo {author} {\bibfnamefont {T.}~\bibnamefont
  {Schmidt}}, \bibinfo {author} {\bibfnamefont {B.}~\bibnamefont {Schmitt}},
  \bibinfo {author} {\bibfnamefont {A.}~\bibnamefont {Streun}}, \bibinfo
  {author} {\bibfnamefont {I.}~\bibnamefont {Vartiainen}}, \bibinfo {author}
  {\bibfnamefont {M.}~\bibnamefont {Vitins}}, \bibinfo {author} {\bibfnamefont
  {X.}~\bibnamefont {Wang}}, \ and\ \bibinfo {author} {\bibfnamefont
  {R.}~\bibnamefont {Wullschleger}},\ }\href@noop {} {\bibfield  {journal}
  {\bibinfo  {journal} {Journal of Synchrotron Radiation}\ }\textbf {\bibinfo
  {volume} {20}},\ \bibinfo {pages} {667} (\bibinfo {year} {2013})}\BibitemShut
  {NoStop}%
\bibitem [{\citenamefont {Bergamaschi}\ \emph {et~al.}(2010)\citenamefont
  {Bergamaschi}, \citenamefont {Cervellino}, \citenamefont {Dinapoli},
  \citenamefont {Gozzo}, \citenamefont {Henrich}, \citenamefont {Johnson},
  \citenamefont {Kraft}, \citenamefont {Mozzanica}, \citenamefont {Schmitt},\
  and\ \citenamefont {Shi}}]{bergamaschi2010mythen}%
  \BibitemOpen
  \bibfield  {author} {\bibinfo {author} {\bibfnamefont {A.}~\bibnamefont
  {Bergamaschi}}, \bibinfo {author} {\bibfnamefont {A.}~\bibnamefont
  {Cervellino}}, \bibinfo {author} {\bibfnamefont {R.}~\bibnamefont
  {Dinapoli}}, \bibinfo {author} {\bibfnamefont {F.}~\bibnamefont {Gozzo}},
  \bibinfo {author} {\bibfnamefont {B.}~\bibnamefont {Henrich}}, \bibinfo
  {author} {\bibfnamefont {I.}~\bibnamefont {Johnson}}, \bibinfo {author}
  {\bibfnamefont {P.}~\bibnamefont {Kraft}}, \bibinfo {author} {\bibfnamefont
  {A.}~\bibnamefont {Mozzanica}}, \bibinfo {author} {\bibfnamefont
  {B.}~\bibnamefont {Schmitt}}, \ and\ \bibinfo {author} {\bibfnamefont
  {X.}~\bibnamefont {Shi}},\ }\href@noop {} {\bibfield  {journal} {\bibinfo
  {journal} {Journal of Synchrotron Radiation}\ }\textbf {\bibinfo {volume}
  {17}},\ \bibinfo {pages} {653} (\bibinfo {year} {2010})}\BibitemShut
  {NoStop}%
\bibitem [{\citenamefont {Guinier}\ \emph {et~al.}(1955)\citenamefont
  {Guinier}, \citenamefont {Fournet},\ and\ \citenamefont
  {Walker}}]{guiner1955small}%
  \BibitemOpen
  \bibfield  {author} {\bibinfo {author} {\bibfnamefont {A.}~\bibnamefont
  {Guinier}}, \bibinfo {author} {\bibfnamefont {G.}~\bibnamefont {Fournet}}, \
  and\ \bibinfo {author} {\bibfnamefont {C.}~\bibnamefont {Walker}},\
  }\href@noop {} {\emph {\bibinfo {title} {Small angle scattering of X-rays}}}\
  (\bibinfo  {publisher} {J. Wiley \& Sons: New York},\ \bibinfo {year}
  {1955})\BibitemShut {NoStop}%
\bibitem [{\citenamefont {Debye}(1915)}]{debyetotalscatt}%
  \BibitemOpen
  \bibfield  {author} {\bibinfo {author} {\bibfnamefont {P.}~\bibnamefont
  {Debye}},\ }\href@noop {} {\bibfield  {journal} {\bibinfo  {journal} {Ann.
  Physik}\ }\textbf {\bibinfo {volume} {46}},\ \bibinfo {pages} {809} (\bibinfo
  {year} {1915})}\BibitemShut {NoStop}%
\end{thebibliography}
%

\end{document}